\documentclass[aps,twocolumn,nofootinbib, superscriptaddress, floatfix,longbibliography]{revtex4-1} 

\usepackage{multirow}
\usepackage{booktabs}
\usepackage{placeins}
\usepackage{soul}
\usepackage{bbm}
\usepackage{amsmath,amssymb,amsbsy,bm,amsfonts,mathrsfs}
\usepackage{graphicx}
\usepackage{wrapfig}
\usepackage{url}
\usepackage{float}
\usepackage{lipsum}
\usepackage{enumitem}
\usepackage{titlesec}
\usepackage{microtype}
\usepackage{slashed}

\usepackage[usenames,dvipsnames,svgnames]{xcolor}
\definecolor{redd}{rgb}{0.8, 0.1,0.2}
\definecolor{navy}{rgb}{0.05, 0.23,0.75}

\usepackage[colorlinks]{hyperref}
\hypersetup{
    colorlinks    =    true,
    citecolor     =    navy,
	linkcolor     =    redd,
	urlcolor      =    navy,
	anchorcolor   =    blue
}

\begin{document}

\title{Dark Matter Freeze-out via Catalyzed Annihilation}

\author{Chuan-Yang Xing}
\email{cyxing@pku.edu.cn}
\affiliation{Department of Physics and State Key Laboratory of Nuclear Physics and Technology, Peking University, Beijing 100871, China}

\author{Shou-hua Zhu}
\email{shzhu@pku.edu.cn}
\affiliation{Department of Physics and State Key Laboratory of Nuclear Physics and Technology, Peking University, Beijing 100871, China}
\affiliation{Collaborative Innovation Center of Quantum Matter, Beijing, 100871, China}
\affiliation{Center for High Energy Physics, Peking University, Beijing 100871, China}

\begin{abstract}

We present a new paradigm of dark matter freeze-out, where the annihilation of dark matter particles is catalyzed. We discuss in detail the regime that the depletion of dark matter proceeds via $2\chi \to 2A'$ and $3A' \to 2\chi$ processes, in which $\chi$ and $A'$ denote dark matter and the catalyst respectively. In this regime, the dark matter number density is depleted polynomially rather than exponentially (Boltzmann suppression) as in classical WIMPs and SIMPs. The paradigm applies for a secluded weakly interacting dark sector with a dark matter in the $\text{MeV-TeV}$ mass range. The catalyzed annihilation paradigm is compatible with CMB and BBN constraints, with enhanced indirect detection signals.

\end{abstract}

\maketitle

\section{Introduction} \label{sec:intro}

The existence of dark matter (DM) is well established with ample evidence from cosmological and astrophysical observations~\cite{Plehn:2017fdg}. Though, the nature of dark matter is still unknown.
To solve this puzzle, tremendous efforts have been devoted to searching for dark matter candidates and studying production mechanisms of dark matter in the early universe.
Among all the mechanisms that reproduce the observed abundance of dark matter, the possibility of thermal dark matter where dark matter keeps in thermal equilibrium with Standard Model (SM) particles in the early universe is especially popular and compelling.

For massive thermal dark matter, DM particles remain in thermal and chemical equilibrium while relativistic. As the universe cools down, DM particles are depleted via some certain processes and the abundance of dark matter goes down.
These processes freeze-out when their interaction rate falls below the expansion rate of the universe, and consequently DM abundance is settled.
There are essentially two kinds of process leading to depletion of DM particles in the literature. The first one is that DM particles annihilate into other particles, mostly SM particles. The other one is via number-changing process of dark matter.
For the former case, the most studied scenario is self-annihilation process~\cite{Lee:1977ua}, e.g. $2\text{DM} \to 2\text{SM}$. Especially, weakly interacting massive particles (WIMPs) that naturally reproduce correct relic abundance attracted extensive attentions~\cite{Bertone:2004pz,Arcadi:2017kky,Roszkowski:2017nbc}.
Other variations on the self-annihilation case include co-annihilation~\cite{Griest:1990kh,Ellis:1998kh,DAgnolo:2018wcn}, semi-annihilation~\cite{DEramo:2010keq} and so on~\cite{Griest:1990kh,DAgnolo:2015ujb,Pospelov:2007mp,ArkaniHamed:2008qn,Feng:2008ya,Belanger:2011ww,Dror:2016rxc,DAgnolo:2017dbv,Garny:2019kua,Maity:2019hre,Kramer:2020oqi,Dey:2016qgf,Cline:2017tka}.
Whereas, for the number-changing process, the most studied process is $3\text{DM} \to 2\text{DM}$ annihilation, dubbed as strongly interaction massive particles (SIMPs)~\cite{Hochberg:2014dra,Hochberg:2014kqa}. Subsequently, other number-changing process are proposed and discussed, including $Z_2$-symmetric SIMPs~\cite{Bernal:2015xba,Bernal:2017mqb}, co-SIMPs~\cite{Smirnov:2020zwf}, etc.

\begin{figure}[t]
    \centering
    \includegraphics[width=8cm]{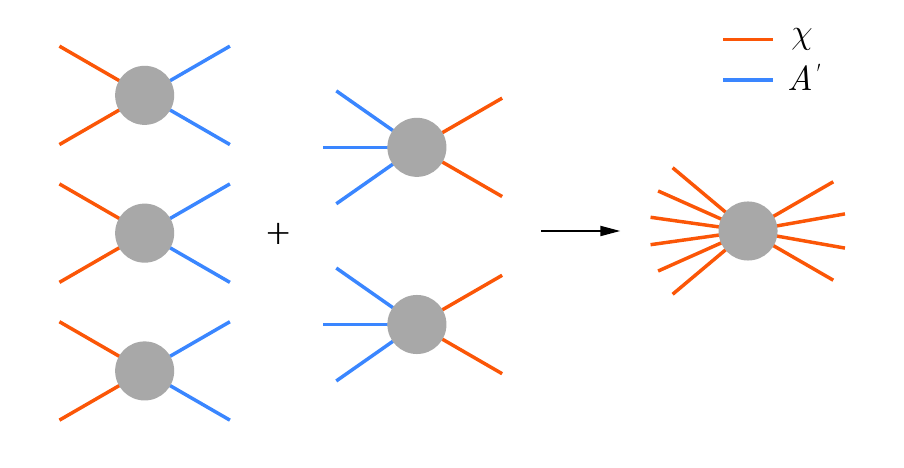}
    \caption{Schematic illustration of the catalyzed annihilation of DM $\chi$ (red line) with a catalyst $A'$ (blue line). Three $2\chi \to 2A'$ processes plus two $3A' \to 2\chi$ effectively deplete the number of DM particles by two.}
    \label{fig:illustration}
\end{figure}

In this Letter, we propose a new pattern of dark matter burning in the early universe beside the two aforementioned kinds of process, where the abundance of dark matter is determined by \emph{catalyzed} processes.
In catalyzed processes, there are some other particles beside DM that act as the catalyst. The catalyst can \emph{enhance the rate of dark matter burning}, yet the catalyst itself is \emph{not consumed in the reaction} \cite{EngelThomas2009Pc,RothenbergGadi2017CCaG,PetrucciRalphH2002Gc:p}. It provides with an alternative reaction pathway to make the reaction happen without changing the reactants and products.
Specifically, in this work we study a simple regime of catalyzed annihilation with two processes leading to depletion of DM particles: $2\chi \to 2A'$ and $3A' \to 2\chi$, where $\chi$ and $A'$ denote dark matter and the catalyst respectively.
We shown in Figure~\ref{fig:illustration} a depiction of how these annihilation channels result in depopulation of DM particles, that is, three $2\chi \to 2A'$ processes together with two $3A' \to 2\chi$ effectively deplete two DM particles.
Note that the assisted annihilation~\cite{Dey:2016qgf,Cline:2017tka,Fitzpatrick:2020vba} are not catalyzed reactions since the assisters are consumed in the reaction. The co-SIMP process $\text{SM}+\chi+\chi \to \text{SM}+\chi$~\cite{Smirnov:2020zwf} is not catalyzed reaction either, since $\chi+\chi \to \chi$ is kinetically forbidden and it's groundless to discuss enhancement of this unphysical process. Same thing happens to Ref.~\cite{Dolgov:2017ujf}.
We acknowledge that catalyzed processes are also considered in the Big Bang Nucleosynthesis (BBN)~\cite{Pospelov:2006sc}.

The observed DM abundance can be reproduced in the catalyzed annihilation paradigm for a wide mass range of dark matter.
We emphasis that the thermal evolution in the catalyzed annihilation paradigm is unique. Different with WIMPs and SIMPs, where DM number density $n_\chi$ tracks Boltzmann distribution and shrinks exponentially before freeze-out, the catalyzed annihilation could lead to a polynomial suppression of $n_\chi$ as the universe cools down,
\begin{equation}
    n_\chi \propto s^{3/2} \propto T^{9/2},
    \label{polynomial_suppression}
\end{equation}
where $s$ and $T$ denote entropy density and temperature of the universe.
Thus, the catalyzed annihilation lasts longer and freezes-out at late times.
To reproduce correct relic abundance, the cross section of DM annihilation $2\chi \to 2A'$ should be enhanced since there is less time to redshift to today~\cite{Dror:2016rxc}, which corresponds to enhanced indirect detection signals.

\section{Catalyzed Freeze-out} \label{sec:catalyzed}

In order for the catalyzed annihilation paradigm to work, there are several requirements listed as follows:
\begin{itemize}
    \item The dark sector is nearly secluded.
    \item The catalyst is long-lived ($\gtrsim 10^{-9}\text{s}$).
    \item The catalyst is slightly lighter than DM ($1 < m_{\chi}/m_{A'} \lesssim 2$).
    \item Annihilation channels as in Figure~\ref{fig:illustration}.
\end{itemize}
The dark sector should be secluded so that the annihilation channels to SM particles freeze-out before the catalyzed annihilation.
The catalyst is lighter than DM and long-lived so that its number density is large and the $3A' \to 2\chi$ process is not suppressed, which ensures that the catalyzed annihilation happens.
If $A'$ decays fast, the paradigm recovers to the secluded DM regime~\cite{Pospelov:2007mp,ArkaniHamed:2008qn}. Besides, the catalyzed annihilation will heat up the dark sector. For simplicity, we assume the dark sector could scatter with SM particles intensely enough to maintain thermal equilibrium with SM particles.

\begin{figure}[t]
    \centering
    \includegraphics[width=8cm]{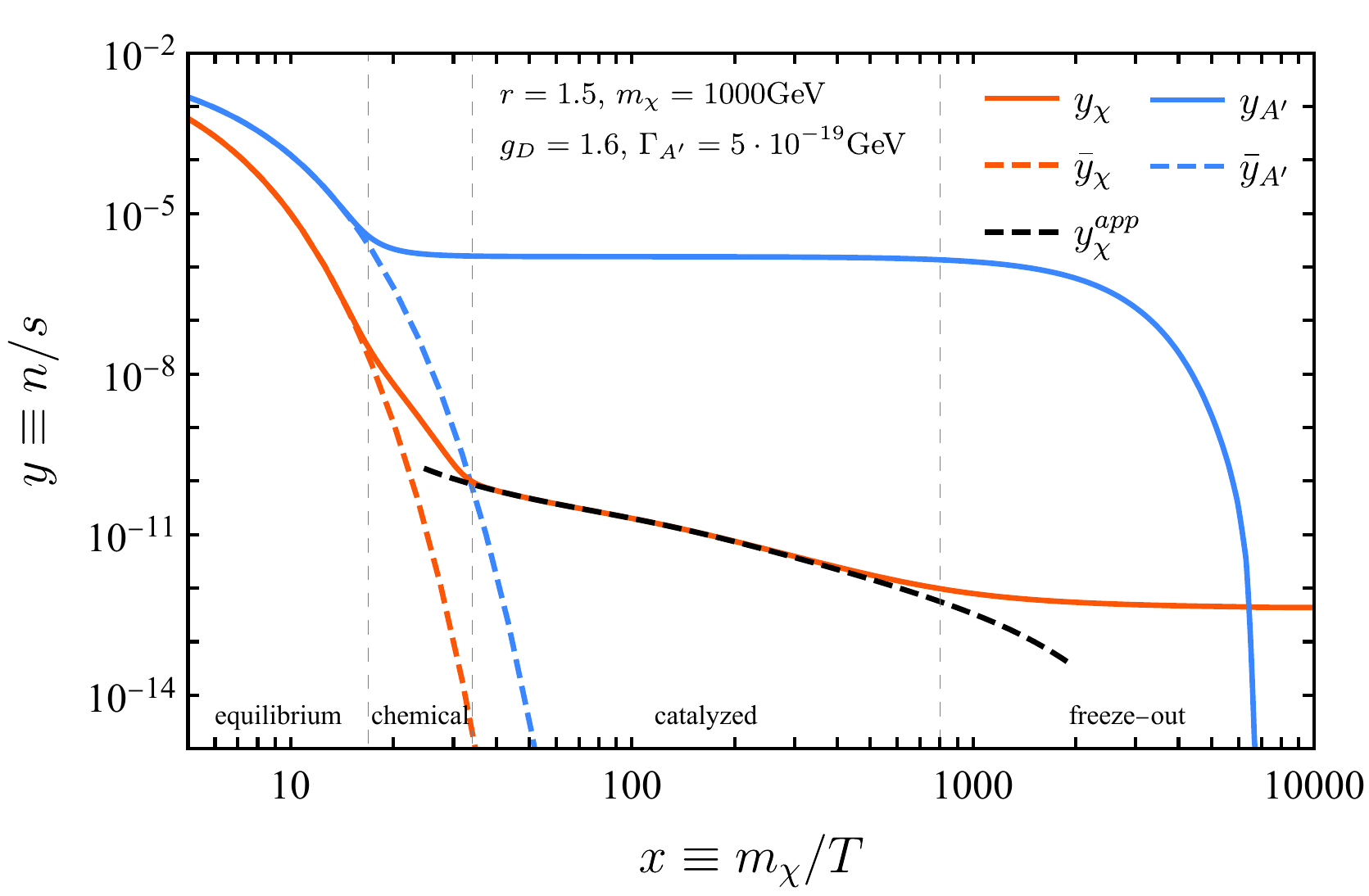}
    \caption{Thermal evolution of DM $\chi$ (solid red) and the catalyst $A'$ (solid blue). The dashed colored curves denote equilibrium yields $\bar{y} \equiv \bar{n}/s$, while the dashed black curve shows the approximate DM yield $y^{app}_\chi \equiv n^{app}_\chi/s$ during the catalyzed annihilation stage with $n_\chi^{app} = \sqrt{   n_{A'}^3 \left< \sigma_3 v^2 \right> / \left< \sigma_2 v \right>  }$ from Eq.~\ref{catalyzed_relation}. The parameters are taken for the model presented in Section~\ref{sec:model}. }
    \label{fig:evolution}
\end{figure}

We show in Figure~\ref{fig:evolution} a typical thermal history of dark matter that freezes-out via catalyzed annihilation. For now, we are focused on the regime that the mass ratio of dark matter and the catalyst $r \equiv m_\chi/m_{A'}$ is no larger than $1.5$. As is shown in the figure, there are four stages in the thermal evolution:
\begin{enumerate}
    \item \emph{Equilibrium stage.} Both $\chi$ and $A'$ stay in chemical equilibrium due to the number-changing processes in the dark sector. The dominate number-changing process is $3A' \to 2\chi$ for $r \gtrsim 1.1$. Other processes with DM in the initial state, e.g. $\chi A'A' \to \chi A'$, are suppressed and negligible, since $n_{\chi} \ll n_{A'}$.
    \begin{equation}
        n_\chi \simeq \bar{n}_\chi, n_{A'} \simeq \bar{n}_{A'}.
    \end{equation}
    $n_{\chi,A'}$ denote the number densities and $\bar{n}_{\chi,A'}$ are the equilibrium densities.
    In the non-relativistic limit, we have $\bar{n}_{\chi,A'} = g_{\chi,A'} \left( \frac{m_{\chi,A'}T}{2\pi} \right)^{3/2} e^{-m_{\chi,A'}/T}$, where $g$ denotes number of internal degrees of freedom.
    \item \emph{Chemical stage.} $\chi$ and $A'$ are chemically decoupled from equilibrium, but they can still maintain chemical equilibrium with each other via the $2\chi \leftrightarrow 2{A'}$ process.
    \begin{equation}
        n_\chi/\bar{n}_\chi \simeq n_{A'}/\bar{n}_{A'}.
    \end{equation}
    \item \emph{Catalyzed annihilation.} 
    As the rate of $2A' \to 2\chi$ (inverse process of $2\chi \to 2A'$) descends exponentially at low temperature, the $3A' \to 2\chi$ process dominates over it. The evolution of DM number density is now controlled by the catalyzed annihilation, i.e. $2\chi \to 2A'$ and $3A' \to 2\chi$. Before freeze-out, the rates of the $2\chi \to 2A'$ and $3A' \to 2\chi$ reactions are much larger than the rate of change of $n_\chi$, as well as the Hubble rate and rates of other reactions (see in Eq.~\ref{boltzmann_equation}). Thus, neglecting the subdominant terms, we get an approximate relation,
    \begin{equation}
        \left< \sigma_2 v \right> n_\chi^2 \simeq \left< \sigma_3 v^2 \right> n_{A'}^3.
        \label{catalyzed_relation}
    \end{equation}
    We used $\left< \sigma_2 v \right>$ and $\left< \sigma_3 v^2 \right>$ to denote the thermally averaged cross sections of $2\chi \to 2A'$ and $3A' \to 2\chi$ respectively. In this stage, since $y_{A'} \equiv n_{A'}/s$ is practically constant and $\left< \sigma_2 v \right>$ and $\left< \sigma_3 v^2 \right>$ are polynomial functions of $T$, Eq.~\ref{catalyzed_relation} indicates that $n_\chi$ is polynomially suppressed. It is similar to the scaling of the number density of the assisting particle after DM freeze-out in Ref.~\cite{Cline:2017tka}.
    \item \emph{Freeze-out.} As the universe expands, the rate of the catalyzed annihilation descends and dark matter freezes-out.
\end{enumerate}

The equilibrium stage ends when the rate of $3A' \to 2\chi$ falls below Hubble constant $H$. The temperature of departure from equilibrium $T_c$ can be determined approximately with,
\begin{equation}
    \left< \sigma_3 v^2 \right> \bar{n}_{A'}^3 \simeq H (\bar{n}_{A'}+\bar{n}_\chi).
    \label{catalyst_freeze-out}
\end{equation}
We note that the annihilation channels to SM particles or the $3A' \to 2A'$ process can also deplete dark sector particles and $T_c$ could be altered if these channels freeze-out later. 
The ending of the chemical stage is insignificant since the freeze-out temperature $T_f$ and relic abundance can be estimated without it.
Lastly, the catalyzed annihilation freezes-out when the rate drops below $H$. $T_f$ is determined by,
\begin{equation}
    \left< \sigma_2 v \right> n_{\chi}^2 \simeq \left< \sigma_3 v^2 \right> n_{A'}^3 \simeq H n_\chi.
    \label{DM_freeze-out}
\end{equation}

The relic abundance of DM can be estimated approximately in the same spirit of WIMPs~\cite{Kolb:1990vq,Dror:2016rxc},
\begin{equation}
    \Omega_\chi = 
    \frac{m_\chi s_0 H_m}{\rho_c s_m}
    \frac{\sqrt{g_{\star,m}}}{\sqrt{g_{\star,f}}} 
    \frac{x_f}{ \left< \sigma_2 v \right> }.
    \label{relic_abundance}
\end{equation}
where $x_f \equiv m_\chi / T_f$.
Since $x_f$ is dependent on $x_c \equiv m_\chi / T_c$, we solve for $x_c$ with Eq.~\ref{catalyst_freeze-out} first. Simplifying Eq.~\ref{catalyst_freeze-out}, we find,
\begin{equation}
    x_c = \frac{r}{2} \log \left[ 0.0024 \frac{ g_{A'}^2 m_\chi^4 M_\text{Pl} \left< \sigma_3 v^2 \right>_c}{g_{\star,c}^{1/2} r^3 x_c} \right],
\label{x_departure_equilibrium}
\end{equation}
where $M_\text{Pl}$ is Planck mass. With $x_c$ determined in Eq.~\ref{x_departure_equilibrium}, we can solve for $n_{A'}$. Substituting the result into Eq.~\ref{DM_freeze-out}, we get,
\begin{equation}
    x_f = \left( 1.2 \frac{g_{\star,f}^2}{g_{\star,c}^{9/4}} \frac{ M_\text{Pl}^{1/2} \left< \sigma_2 v \right>_f \left< \sigma_3 v^2 \right>_f }{ m_\chi \left< \sigma_3 v^2 \right>_c^{3/2} } x_c^6 \right)^{1/5}.
\label{x_freeze-out}
\end{equation}
The subscripts $m,c,f$ in the equations above mark the temperatures, $T=m_\chi, T_c, T_f$, respectively, for the quantities, including entropy density $s$, Hubble constant $H$, effective degrees of freedom $g_\star$\footnote{We neglect the differences between effective entropy degrees of freedom $g_{\star,s}$ and effective energy degrees of freedom $g_\star$ as in Ref.~\cite{Dror:2016rxc}} and the thermally averaged cross sections.
Note that if $x_c$ is delayed due to annihilation to SM particles or $3A' \to 2A'$, Eq.~\ref{x_freeze-out} and Eq.~\ref{x_departure_equilibrium} should be modified to include these processes and DM will freeze-out earlier in this case with a smaller relic abundance.

Based on the partial wave unitarity limit~\cite{Griest:1989wd}, $\sigma_2 v \leq \frac{4\pi}{m_\chi^2 v}$, we can estimate the upper bound of DM mass from Eq.~\ref{relic_abundance} for the catalyzed annihilation paradigm. With $x_f \gtrsim 100$, we deduce $m_\chi \lesssim 100 \text{TeV}.$
Compared to SIMP dark matter that lives in the $\text{MeV}$ scale~\cite{Hochberg:2014dra}, it is compelling to notice that $3 \to 2$ process can apply to such a heavy dark matter.

In order to study the thermal evolution and DM freeze-out in a quantitative way, we turn to the Boltzmann equations. As is discussed above, we neglect the subdominant $3\to2$ annihilation channels, including $\chi A' A' \to \chi A'$, $\chi \chi A' \to A' A'$, $\chi \chi A' \to \chi \chi$, $\chi \chi \chi \to \chi A'$ and assume $3A' \to 2A'$ is subdominant. If $A'$ decays to SM particles, the Boltzmann equations reads,
\begin{align}
\label{boltzmann_equation}
\dot{n}_\chi  &+  3H n_\chi       =      - \left< \sigma_2 v \right> \left( n_{\chi}^2  -  \bar{n}_{\chi}^2  \frac{n_{A'}^2}{\bar{n}_{A'}^2} \right)       \\ 
&+ \left< \sigma_3 v^2 \right> \left( n_{A'}^3  -  \bar{n}_{A'}^3  \frac{n_\chi^2}{\bar{n}_\chi^2} \right) ,        \nonumber \\
\dot{n}_{A'}   &+  3H n_{A'}       =  + \left< \sigma_2 v \right> \left( n_{\chi}^2  -  \bar{n}_{\chi}^2  \frac{n_{A'}^2}{\bar{n}_{A'}^2} \right)       \nonumber \\
&-\frac{3}{2} \left< \sigma_3 v^2 \right> \left( n_{A'}^3  -  \bar{n}_{A'}^3  \frac{n_\chi^2}{\bar{n}_\chi^2} \right)       -      
\left< \Gamma_{A'} \right> (n_{A'} - \bar{n}_{A'}).     \nonumber
\end{align}
The yield $y_{\chi,A'} \equiv n_{\chi,A'}/s$ can be solved numerically and are shown in Figure~\ref{fig:evolution}.

\section{Mass Ratio} \label{sec:mass_ratio}

\begin{figure}[t]
    \centering
    \includegraphics[width=8cm]{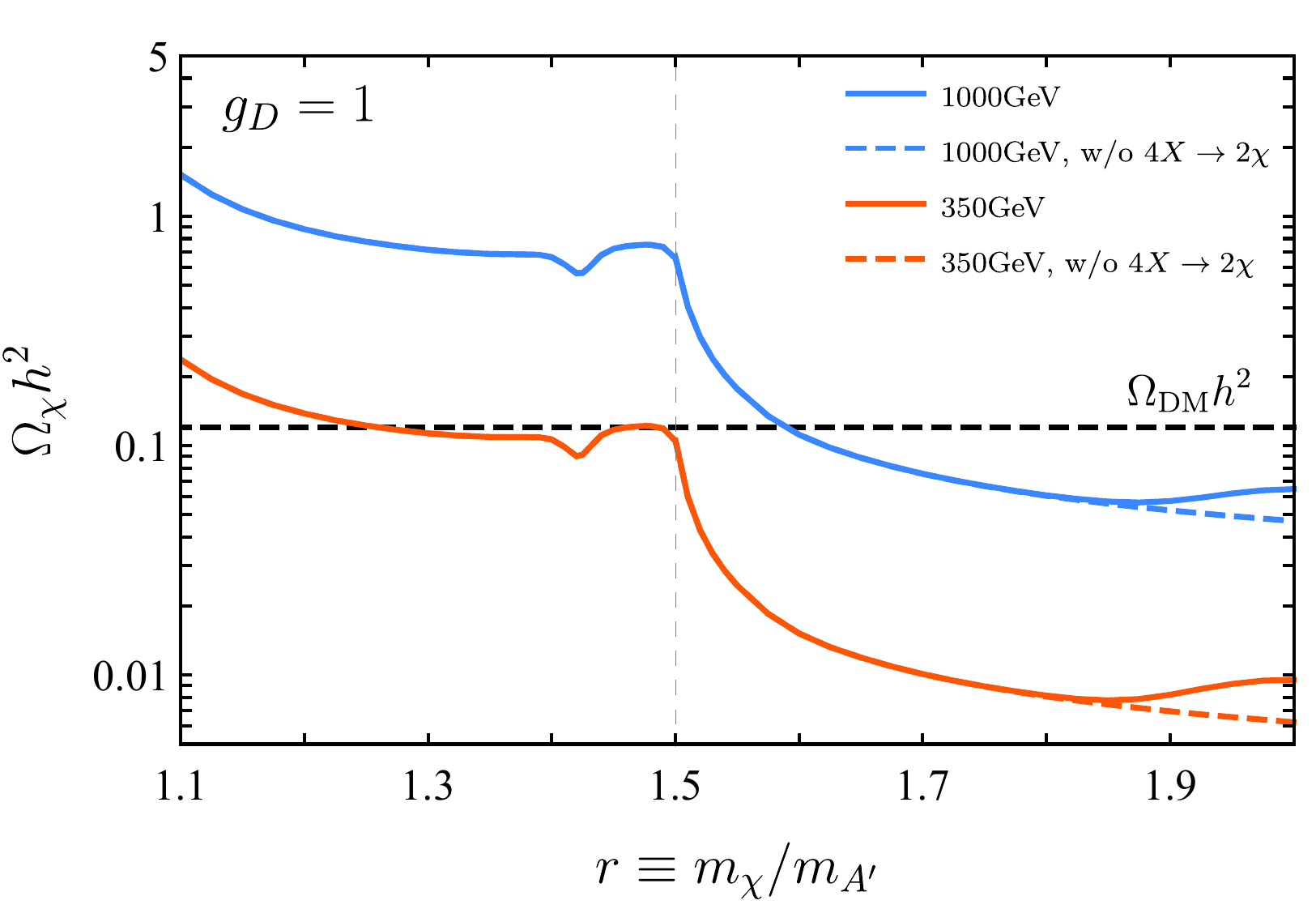}
    \caption{Curves of relic abundance $\Omega_\chi h^2$ for $m_\chi=350\text{GeV}$ (solid red) and $m_\chi=1000\text{GeV}$ (solid blue) with respect to different mass ratio $r$. For dashed colored curves, the $4A' \to 2\chi$ process is neglected. The dashed black curve denotes the observed DM relic abundance~\cite{Aghanim:2018eyx}. The parameters are taken for the model presented in Section~\ref{sec:model}.}
    \label{fig:mass_ratio}
\end{figure}

In previous section, we concentrated on the mass ratio $r \leq 1.5$. In fact, the catalyzed annihilation paradigm can go beyond this limit. Firstly, when the mass ratio is slightly larger than $1.5$, i.e. $3m_{A'} < 2m_\chi$, $\left< \sigma_3 v^2 \right>$ is exponentially suppressed as the temperature goes down,
\begin{equation}
    \left< \sigma_3 v^2 \right> \propto e^{-(2r-3)x/r},
    \label{v3s_for_r>1.5}
\end{equation}
where $x \equiv m_\chi/T$. During the catalyzed annihilation period, with less DM particles produced via $3A' \to 2\chi$ process since the cross section is smaller, the DM number density shrinks more sharply. Consequently, the catalyzed annihilation freezes-out much earlier.

As the mass ratio grows, when $r\lesssim2$, it is intriguing to notice that the $4A' \to 2\chi$ process may play a part in the catalyzed annihilation.
To be specific, after a period of catalyzed annihilation governed by $2\chi \to 2A'$ and $3A' \to 2\chi$ as usual, there would be an extra stage of catalyzed annihilation predominated by $2\chi \to 2A'$ and $4A' \to 2\chi$, in which the non-suppressed $4A' \to 2\chi$ process takes over the role of converting $A'$ to DM particles since the cross section of $3A' \to 2\chi$ is exponentially suppressed (Eq.~\ref{v3s_for_r>1.5}), 
Similar to Eq.~\ref{catalyzed_relation}, we can deduce an approximate relation that holds in this stage,
\begin{equation}
    \left< \sigma_2 v \right> n_\chi^2 \simeq \left< \sigma_4 v^3 \right> n_{A'}^4,
\end{equation}
where $\left< \sigma_4 v^3 \right>$ denotes the thermally averaged cross section for $4A' \to 2\chi$.
The presence of $4A' \to 2\chi$ is essential. If it is neglected, as is discussed previously, $n_\chi$ shrinks sharply and dark matter freezes-out early. Once $4A' \to 2\chi$ takes charge, the sharply falling of $n_\chi$ is bent and the polynomial suppression recovers (compared to Eq.~\ref{polynomial_suppression}).
\begin{equation}
    n_\chi \propto  s^{2} \propto T^{6}.
\end{equation}
Thus, the catalyzed annihilation freezes-out at later times, leading to enhanced DM relic abundance.

For even larger mass ratio, we expect the processes with more catalysts annihilating to two DM particles, e.g. $5A' \to 2\chi$, to possibly play a role in the catalyzed annihilation, especially when the dark sector is strongly coupled.

We show in Figure~\ref{fig:mass_ratio} the variation of DM relic abundance $\Omega_\chi h^2$ with different mass ratio. When the mass ratio passes the critical value of $1.5$, $\Omega_\chi h^2$ decreases rapidly. On the other hand, for $r\lesssim2$, relic abundance is uplifted if $4A' \to 2\chi$ process is included.

\section{A Model} \label{sec:model}

The requirements for realization of the catalyzed annihilation presented in Section~\ref{sec:catalyzed} can be easily met in many models.
In this section, we simply present a dark photon model~\cite{Holdom:1985ag,Raggi:2015yfk,Bauer:2018onh,Lao:2020inc,Fabbrichesi:2020wbt} with a Dirac fermion $\chi$ charged under a novel $U(1)'$ gauge group and $A'$ being the gauge field.
The Lagrangian for the dark sector is,
\begin{equation}
    \mathcal{L}_\text{DS}  = 
    -\frac{1}{4} F'_{\mu \nu} F'^{\mu \nu} +
    \frac{1}{2} m_{A'}^2 A'_\mu A'^\mu +
    \bar{\chi} (i \slashed{D} - m_\chi) \chi,
\end{equation}
where $\slashed{D}=\slashed{\partial}-i g_D \slashed{A'}$ and $g_D$ is the gauge coupling constant. The mass of the dark photon can be generated via the Higgs mechanism (or Stueckelberg mechanism~\cite{Stueckelberg:1900zz,Ruegg:2003ps}). We assume the dark Higgs boson is heavy and can be neglected.
SM particles are neutral under the $U(1)'$ gauge group. The dark photon can be kinetically mixed with SM hypercharge field.
\begin{equation}
    \mathcal{L}_\text{mix} = -\frac{\epsilon}{2 \cos{\theta_W}} F'_{\mu \nu}B^{\mu \nu}.
    \label{kinetic_mixing}
\end{equation}
$\epsilon$ is the mixing constant and $\theta_W$ denotes the Weinberg angle. $B^{\mu}$ is SM hypercharge field. Therefore, the dark sector can communicate with SM particles via the mixing and the dark photon $A'$ can decay to SM particles. $\epsilon$ should be small so that the dark photon is long-lived and acts as the catalyst. 
Additionally, the kinetic mixing could not keep the dark sector in thermal equilibrium with SM since $\epsilon$ is small. In order to thermalize the dark sector, we need another portal for the dark sector to interact with SM particles, which might be the dark Higgs. Anyhow, we won't model this part and simply assume that the dark sector stays in thermal equilibrium before freeze-out.

\begin{figure}[t]
    \centering
    \includegraphics[width=8cm]{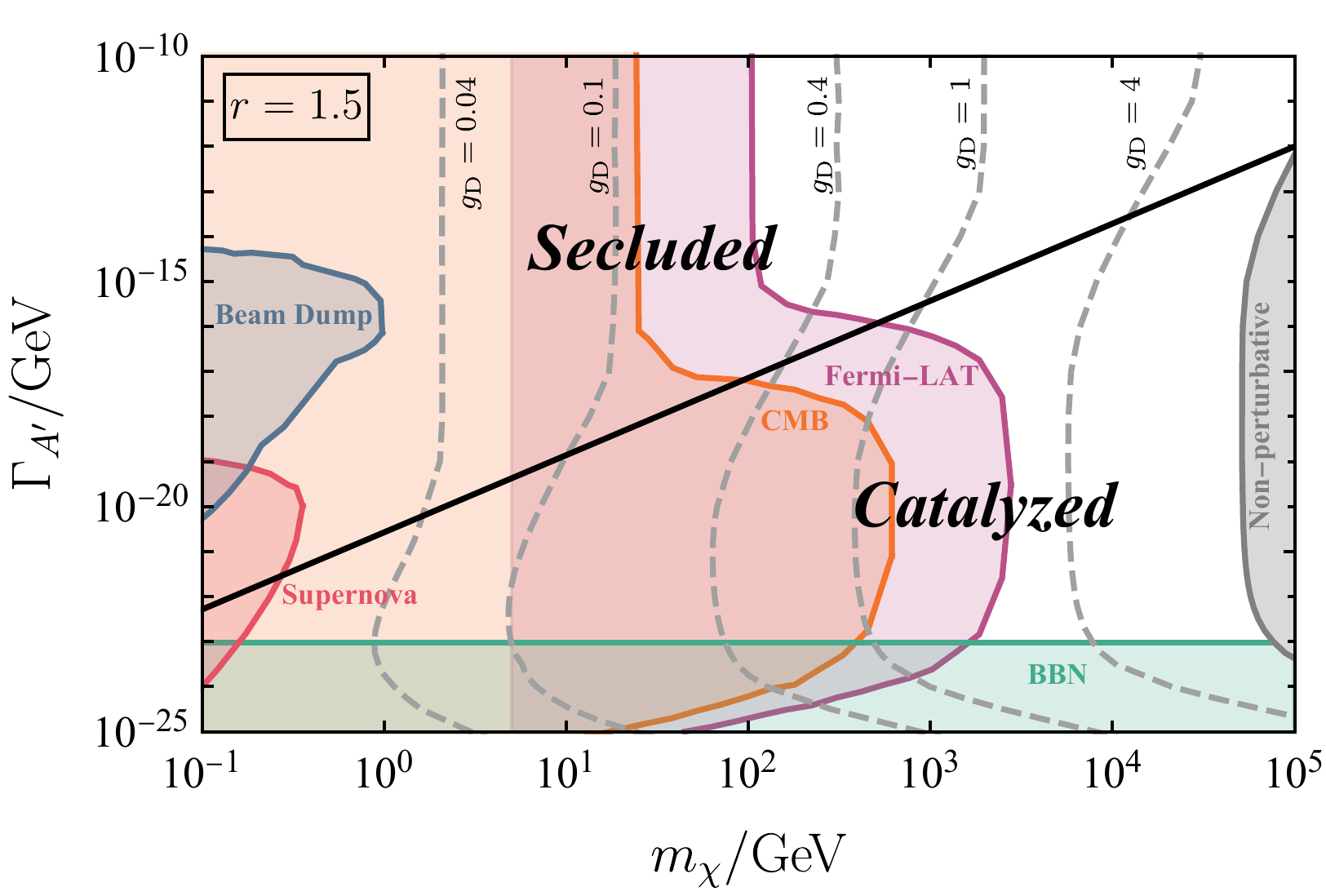}
    \caption{Phase diagram and constraints for the dark photon model in the $(m_\chi, \Gamma_{A'})$ plane with $r=1.5$. The solid black curve marks the boundary of the secluded phase and the catalyzed phase of the model. Correct relic abundance can be reproduced for each point in the figure by varying the value of $g_D$. Especially, we show in gray dashed curves for five different values of $g_D$ that reproduce correct relic abundance. The non-perturbative region is painted gray, while the color-shaded regions denote the bounds from various experiments and observations correspondingly.}
    \label{fig:phase}
\end{figure}

We show in Figure~\ref{fig:phase} different phases for the model in the calculation of DM relic abundance. For short-lived dark photon, before dark matter freezes-out, it simply stays in equilibrium with SM particles via the decay and inverse-decay process. When DM particles annihilates into the dark photon, it immediately decays. This is the secluded phase of the model.
On the other hand, when the dark photon width $\Gamma_{A'}$ is small, the catalyzed annihilation emerges. 
It is a continuous shift, since the decay of the dark photon can occur during the catalyzed annihilation. 
When the dark photon decays after DM freeze-out, $\Omega_\chi h^2$ is independent with $\Gamma_{A'}$.
If $A'$ is sufficiently long-lived ($\Gamma_{A'} \lesssim 10^{-23} \text{GeV}$), it would come to dominate the energy density of the universe. When it decays, considerable entropy is produced and DM abundance is diluted. This effect can help to circumvent the upper bound of DM mass~\cite{Gelmini:2006pq,Dev:2016xcp,Berlin:2016vnh} (see in Figure~\ref{fig:phase}).
We show in dashed gray curves for five different values of $g_D$ that reproduce the observed relic abundance in Figure~\ref{fig:phase}.

The catalyzed annihilation paradigm is constrained by numerous terrestrial and celestial experiments and observations. Firstly, the residual annihilation of $2\chi \to 2A'$ after freeze-out will distort the anisotropy of the Cosmic Microwave Background (CMB) since the decay products of $A'$ are electrically charged particles~\cite{Padmanabhan:2005es, Galli:2009zc, Kawasaki:2015peu, Slatyer:2015jla, Liu:2016cnk, Aghanim:2019ame, Cang:2020exa}. 
Similarly, the signal of DM annihilation at present is detectable in indirect detection experiments~\cite{Adriani:2013uda, Ackermann:2015zua, Fermi-LAT:2016uux, Aguilar:2016kjl, Profumo:2017obk, Abdallah:2018qtu}.
The signal is enhanced compared to WIMPs with $\left< \sigma_2 v \right>_0 \simeq 5 \cdot 10^{-25} \text{cm}^{3} \text{s}^{-1}$ for $s$-wave annihilation.
We used bounds from Fermi-LAT experiment to constrain our model in Figure~\ref{fig:phase}.
The late time decay of $A'$, on the other hand, is also stringently constrained by CMB~\cite{Chen:2003gz, Acharya:2019owx, Acharya:2019uba} as well as BBN~\cite{Kawasaki:2004qu, Jedamzik:2006xz, Kawasaki:2017bqm}.
We note that these bounds can be evaded in models that the catalyst decays into neutrinos or dark radiations~\cite{Ichiki:2004vi, Poulin:2016nat, Nygaard:2020sow}.
Since the dark sector is highly secluded, the constraints on scattering with nucleons~\cite{Cui:2017nnn,Aprile:2018dbl,Leane:2020wob,Leane:2021ihh} are evaded.

For light dark photon, beam dump and fixed target experiments provide great sensitivity on the mixing coupling constant $\epsilon$~\cite{Bergsma:1985is, Bjorken:1988as, Davier:1989wz, Blumlein:2013cua}. There are also lots of new experiments~\cite{Anelli:2015pba, Feng:2017uoz, Berlin:2018pwi, Adrian:2018scb} proposed in recent years that are focused on long-lived particles.
Besides, the long-lived dark photon can enhance the cooling of supernova and the constraints from SN 1987A~\cite{Chang:2016ntp, Chang:2018rso, Sung:2019xie} is widely discussed. These bounds on the dark photon model are considered and presented in Figure~\ref{fig:phase}.

\section{Conclusion and discussion}

We proposed a novel paradigm for thermal relic dark matter, yielding the observed relic abundance. The distinctive wisdom of the paradigm is that the dark matter freeze-out proceeds via catalyzed annihilation. We discussed in detail the scenario that the catalyzed annihilation includes $2\chi \to 2A'$ and $3A' \to 2\chi$, where $\chi$ and $A'$ are dark matter and the catalyst respectively. The paradigm applies for a wide mass range of dark matter, from $\text{1MeV}$ to $\text{100TeV}$, with a unique thermal history compared with WIMPs and SIMPs. Besides, the paradigm offers rich phenomenology including indirect DM search and long-lived particles.

We note that thermal decoupling effects can significantly modify dark matter relic abundance~\cite{Kuflik:2015isi, Pappadopulo:2016pkp, Farina:2016llk, DAgnolo:2017dbv, Kuflik:2017iqs, Fitzpatrick:2020vba}. We leave this to future works~\cite{Xing:inprep}.
Additionally, catalyzed annihilation dominating DM abundance can go far beyond the reactions considered here and should be investigated further.

\section*{Acknowledgments}

C.Y.X. would like to thank Yan-Fang Bai for encouragement.
This work is supported by the National Science Foundation of China under Grants No. 11635001, 11875072.

\bibliography{catalyzed}

\end{document}